\documentclass[12pt]{article}
\usepackage{indentfirst}
\usepackage{amsfonts}
\usepackage{epsf}
\usepackage{graphicx}
\usepackage{subfigure}
\usepackage{psfrag}
\newcommand{\double}{\baselineskip 2.5 \baselineskip}

\title{Tunneling effect of the spin-2 Bose condensate driven by external
magnetic fields }
\author{\begin{tabular}{c}{\small Zhao-xian Yu$^1$, Zhi-yong Jiao$^{1,2}$}  \\
{\small 1. Department of Applied Physics, University of Petroleum
(East
China), }\\
{\small Dongying 257061, Shandong Province, P.R.China}\\{\small 2.
Photonics Center, College of Physics, Nankai University, Tianjin
300071, P.R.China}\end{tabular}}

\date{}

\begin{document}
\maketitle \double
\begin{minipage}[h]{140mm}
\abstract{In this paper, we have studied tunneling effect of the
spin-2 Bose condensate driven by external  magnetic field.   We
find that the population transfers among spin-0 and  spin-$\pm1$,
spin-0 and spin-$\pm2$ exhibit the step structure under the
external cosinusoidal magnetic field respectively, but there do
not exist step structure among spin-$\pm1$ and spin-$\pm2$.  The
tunneling current among spin-$\pm1$  and spin-$\pm2$ may exhibit
periodically oscillation behavior, but among spin-0 and
spin-$\pm1$, spin-0 and spin-$\pm2$, the tunneling currents
exhibit irregular oscillation behavior.
}
\end{minipage}

\section{Introduction}

Recent advance of experimental techniques on Bose-Einstein
condensate (BEC) prompts us to closely and seriously look into
theoretical possibilities which were mere imagination for
theoreticians in this field. This is particularly true for spinor
BEC where all hyperfine states of an atom Bose-condensed
simultaneously, keeping these "spin" states degenerate and active.
Recently, Barrett et al [1] have succeeded in cooling $Rb^{87}$
with the hyperfine state $F=1$  by all optical methods without
resorting to a usual magnetic trap in which the internal degrees
of freedom is frozen. Since the spin interaction of the $Rb^{87}$
atomic system is ferromagnetic, based on the refined calculation
of the atomic interaction parameters by Klausen et al [2], we now
obtain concrete examples of the three-component spinor BEC ($F=1$,
$m_{F}=1,0,-1$) for both antiferromagnetic ($Na^{23}$) [3] and
ferromagnetic interaction cases. In the present spinor BEC the
degenerate internal degrees of  freedom play an essential role to
determine the fundamental physical properties. There is a rich
variety of topological defect structures, which are already
predicted in the earlier studies [4,5] on the spinor BEC. Law et
al [6] constructed an excellent algebraic representation of the
$F=1$ BEC Hamiltonian to study the exact many-body states, and
found that spin-exchange interactions cause a set of collective
dynamic behavior of BEC. Since the spinor BEC appears feasible by
using the $F=2$ multiplet of bosons, it is necessary to
investigate the ground-state structure and magnetic response of
$F=2$  spinor BEC. Recently, Ciobanu et al [7] generalized the
approach for the $F=1$ spinor BEC to study the ground state
structure of the $F=2$ spinor BEC. They found that there are three
possible phases in zero magnetic field, which are characterized by
a pair of parameters describing the ferromagnetic order and the
formation of singlet pairs. From current estimates of scattering
lengths, they also found that the spinor BEC's of $Rb^{87}$ and
$Na^{23}$ have a polar ground state, whereas those of $Rb^{85}$
and $Rb^{83}$ are cyclic and ferromagnetic, respectively. Koashi
et al [8] studied the  exact eigenspectra and eigenstates of $F=2$
spinor BEC. They found that, compared to $F=1$ spinor BEC, the
$F=2$ spinor BEC exhibits an even richer magnetic response due to
quantum correlations among three bosons. Recently, Zhang et al [9]
studied dynamic response of the $F=2$ spinor BEC under the
influence of external magnetic fields, they found that when the
frequency and the reduced amplitude of the longitudinal magnetic
field are related in a specific manner, the population of the
initial spin-0 state will be dynamically localized during time
evolution. In this paper, we shall investigate the tunneling
effect of the spin-2 Bose condensate driven by external magnetic
fields.

\section{ Model}

We consider the $F=2$ spinor BEC subject to a spatial weak uniform
magnetic fields which  consist of longitudinal and transverse
components. Without loss of generality, the transverse direction
of the field is chosen to be along the $x$ axis, i.e.,
$\hat{B}(t)=B_l(t)\hat{z}+B_x\hat{x}$. In such a case, the
second-quantized Hamiltonian of the system is [9]
\begin{equation}
H=H_0+H_B,
\end{equation}
\begin{equation}
H_0=\frac{c_1}{2}\hat{F}\cdot\hat{F}+\frac{2c_2}{5}\hat{S}_+\hat{S}_-,
\end{equation}
$$
H_B=-\mu_Bg_fB_l(t)(\hat{a}_2^\dag\hat{a}_2+\hat{a}_1^\dag\hat{a}_1-
\hat{a}_{-1}^\dag\hat{a}_{-1}-\hat{a}_{-2}^\dag\hat{a}_{-2})$$
\begin{equation}-\mu_Bg_fB_x(\hat{a}_2^\dag\hat{a}_1+\sqrt{\frac{3}{2}}\hat{a}_1^\dag\hat{a}_0+
\sqrt{\frac{3}{2}}\hat{a}_{0}^\dag\hat{a}_{-1}+\hat{a}_{-1}^\dag\hat{a}_{-2}+H.c.),
\end{equation}
here, the $5\times5$ spin matrices $
\hat{F}_i=\hat{a}_\alpha^\dag(F_i)_{\alpha\beta}\hat{a}_\beta
 (i=x, y, z),
\hat{S}_+=\hat{S}_-^\dag=(\hat{a}_0^\dag)^2/2-\hat{a}_1^\dag\hat{a}_{-1}^\dag+\hat{a}_2^\dag\hat{a}_{-2}^\dag$,
and $c_i=(\bar{c_i}\int d\vec{r}|\phi|^4)$. $\bar{c_0}$,
$\bar{c_1}$, and $\bar{c_2}$ are related to scattering lengths
$a_0$, $a_2$, and $a_4$ of the two colliding bosons.

Similar to Ciobanu et al [7], mean-field approximation is used
such that the field operators $\hat{a}_\alpha$ are replaced by $c$
numbers $a_\alpha=\sqrt{P_\alpha}e^{i\theta_\alpha}$, where
$P_\alpha=N_\alpha/N$ is the population in spin $\alpha$, and
$\theta_\alpha$ the phase of wave function $a_\alpha$.
Furthermore, since this paper deals with the quantum coherent
behavior of the system under the influence of the external
magnetic fields, we assume that the initial spin state of the BEC
is the eigenstate in the absence of external fields, then
contribution from Hamiltonian (2) is a constant energy shift and
can be neglected in the study of the dynamics. The semiclassical
equations of motion in Heisenberg representation can be derived
from the Hamiltonian $H_B$ (here, the state vector
$a=(a_2,...,a_{-2})^T $ is introduced)
\begin{equation}
i\dot{a}=H_{eff}(t)a,
\end{equation}
where
\begin{equation}
H_{eff}(t)=-b_l(t)F_{z}-b_xF_x.
\end{equation}
Here $b_l(t)=\mu_Bg_fB_l(t)$ and $b_x=\mu_Bg_fB_x.$ In this paper,
we consider the case that the system begins with unperturbed
spin-0 state $a(0)=(0,0,1,0,0)^T$. If considering the case that
the transverse magnetic field $b_x$ is weak, one can get the time
evolution of the population of spin-$\alpha$ state as [9]
\begin{equation}
P_\alpha(t)=|a_\alpha(t)|^2=|\sum_{\beta=-2}^{2}d_{\beta\alpha}^2(\pi/2)d_{\beta0}^2
(\pi/2)exp(-i\alpha\int_0^tb_l(\tau)d\tau)e^{i\beta\lambda}|^2,
\end{equation}
where
\begin{equation}
\lambda=\sqrt{[\int_0^tb_x\cos[\int_0^\tau
b_l(\tau)d\tau]d\tau]^2+[\int_0^tb_x\sin[\int_0^\tau
b_l(\tau)d\tau]d\tau]^2},
\end{equation}
\begin{equation}
d_{\beta\alpha}^2(\theta)=<2\beta|e^{-i\theta
F_y}|2\alpha>
(\beta,\alpha=2,...,-2).
 \end{equation}

From figures 1-3, we find that  the population transfers among
spin-0 and  spin-$\pm1$, spin-0 and spin-$\pm2$ exhibit the step
structure under the external cosinusoidal magnetic field
$b_l(t)=b\cos(\omega t)$ respectively, but there do not exist step
structure among spin-$\pm1$ and spin-$\pm2$.

\begin{figure}[hpbt]
\includegraphics[width=0.7\textwidth,angle=0]{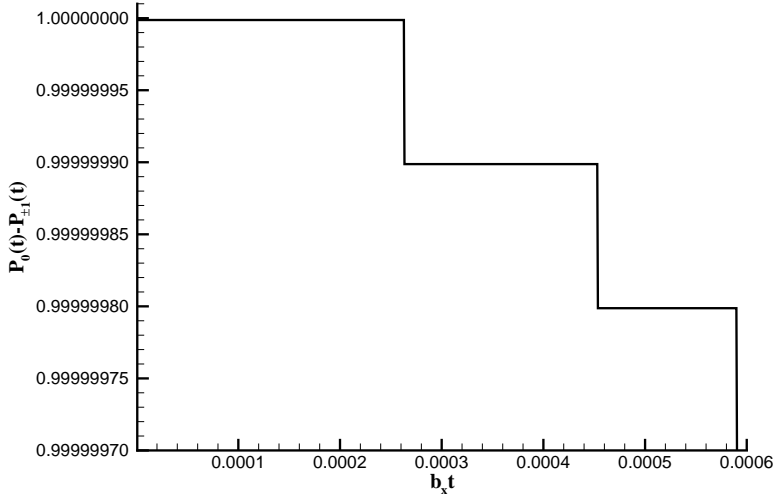}
\caption{Time evolution of population transfer  $P_0(t)-P_{\pm
1}(t)$ with an external cosinusoidal  magnetic field,
$b_x/\omega=0.4, b/\omega=0.8$.}
\end{figure}

\begin{figure}[hpbt]
\includegraphics[width=0.7\textwidth,angle=0]{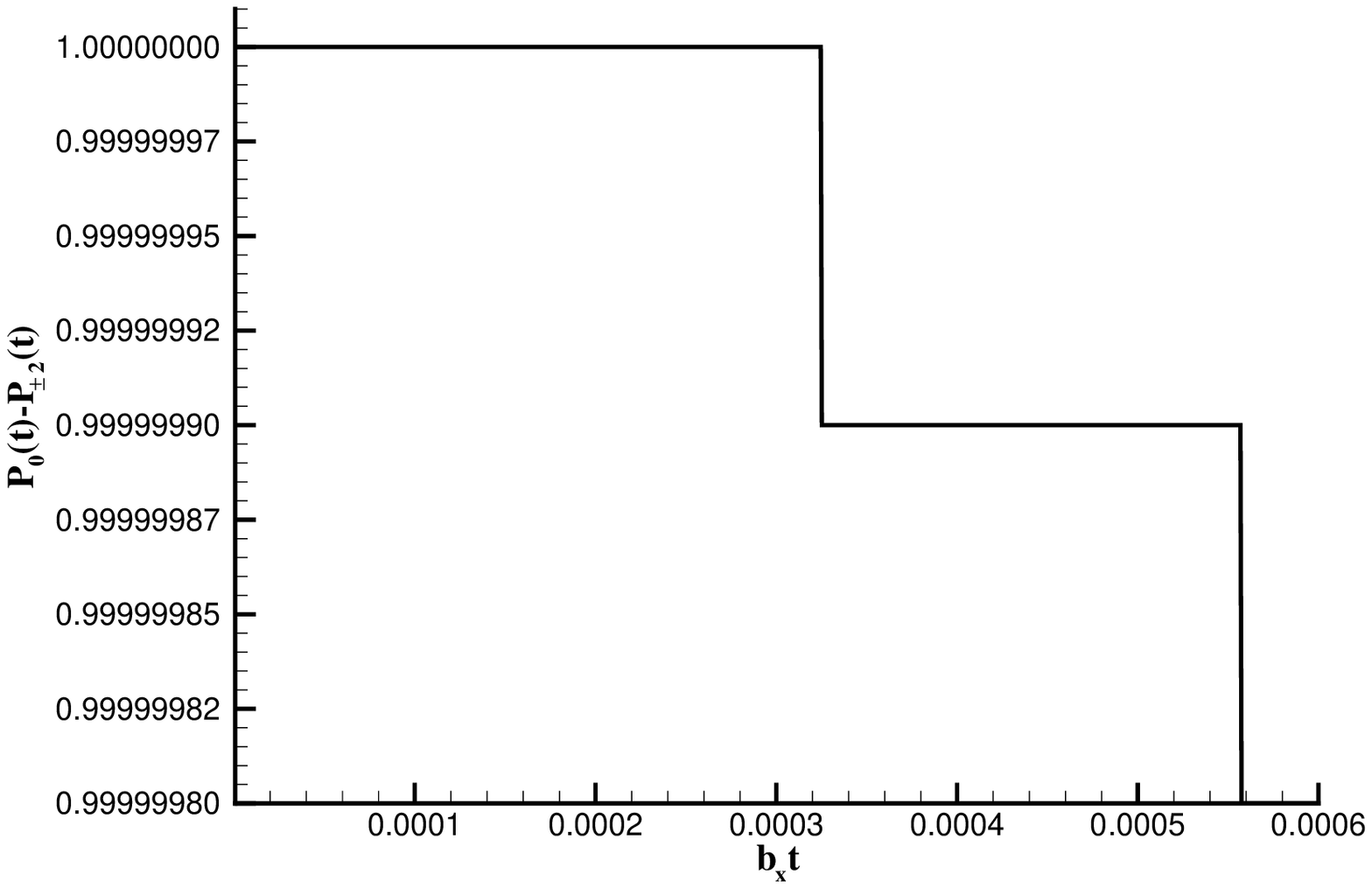}
\caption{Time evolution of population transfer $P_0(t)-P_{\pm
2}(t)$ with an external cosinusoidal magnetic field,
$b_x/\omega=0.4, b/\omega=0.8$.}
\end{figure}

\begin{figure}[hpbt]
\includegraphics[width=0.7\textwidth,angle=0]{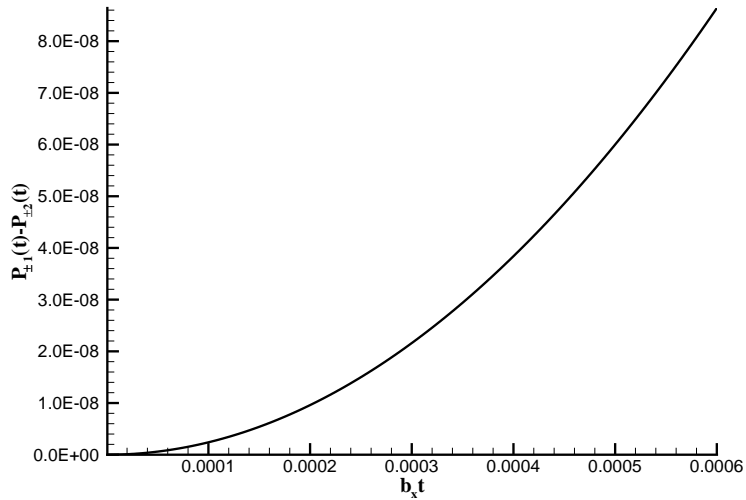}
\caption{Time evolution of population transfer $P_{\pm
1}(t)-P_{\pm 2}(t)$ with an external cosinusoidal magnetic
field,$b_x/\omega=0.4, b/\omega=0.8$.}
\end{figure}

\section{ Tunneling effect of the spin-2 Bose condensate driven by external
 magnetic fields }
In this section, we shall study the tunneling effect between two
different spin components. The tunneling currents between
spin-$\alpha$ and spin-$\beta$
$(\alpha\neq\beta,\alpha,\beta=-2,-1,0,1,2)$ can be defined by
\begin{equation}
I_{\alpha\rightarrow\beta}(t)=\frac{d}{dt}(P_\alpha-P_\beta),
\end{equation}
It is easy to get
\begin{equation}
I_{0\rightarrow\pm1}(t)=(3-\frac{15}{2}\cos^2\lambda)\sin(2\lambda)\frac{d\lambda
}{dt},
\end{equation}
\begin{equation}
I_{0\rightarrow\pm2}(t)=(\frac{3}{4}-\frac{15}{4}\cos^2\lambda)\sin(2\lambda)\frac{d\lambda
}{dt},
\end{equation}
\begin{equation}
I_{\pm1\rightarrow\pm2}(t)=\frac{3}{8}(10\cos^2\lambda-6)\sin(2\lambda)\frac{d\lambda
}{dt},
\end{equation}
where $\lambda$ is determined by Eq. (7).
\\
Figures 4-6 give the evolutions of the tunneling current with the
external  magnetic field $b_l(t)=b\cos(\omega t)$.  The tunneling
current among spin-$\pm1$ and spin-$\pm2$ may exhibit periodically
oscillation behavior, but among spin-0 and spin-$\pm1$, spin-0 and
spin-$\pm2$, the tunneling currents exhibit irregular oscillation
behavior.

\begin{figure}[h]
\includegraphics[width=0.7\textwidth,angle=0]{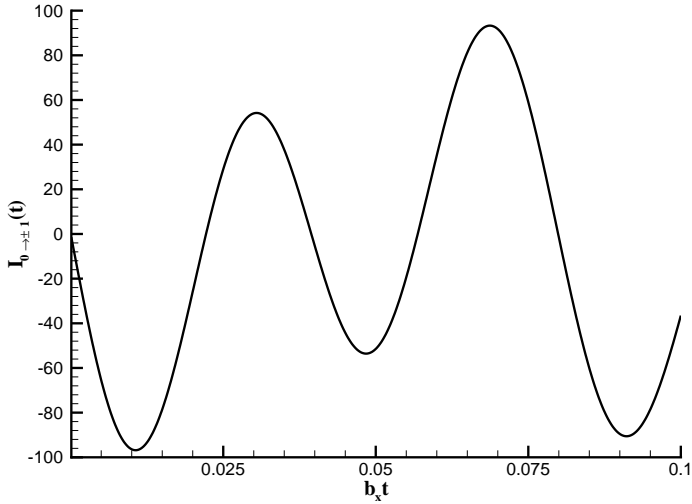}
\caption{Time evolution of tunneling current $I_{0\rightarrow\pm
1}(t)$ with an external cosinusoidal magnetic field,
$b_x/\omega=40, b/\omega=8$.}
\end{figure}

\begin{figure}[htbp]
\includegraphics[width=0.7\textwidth,angle=0]{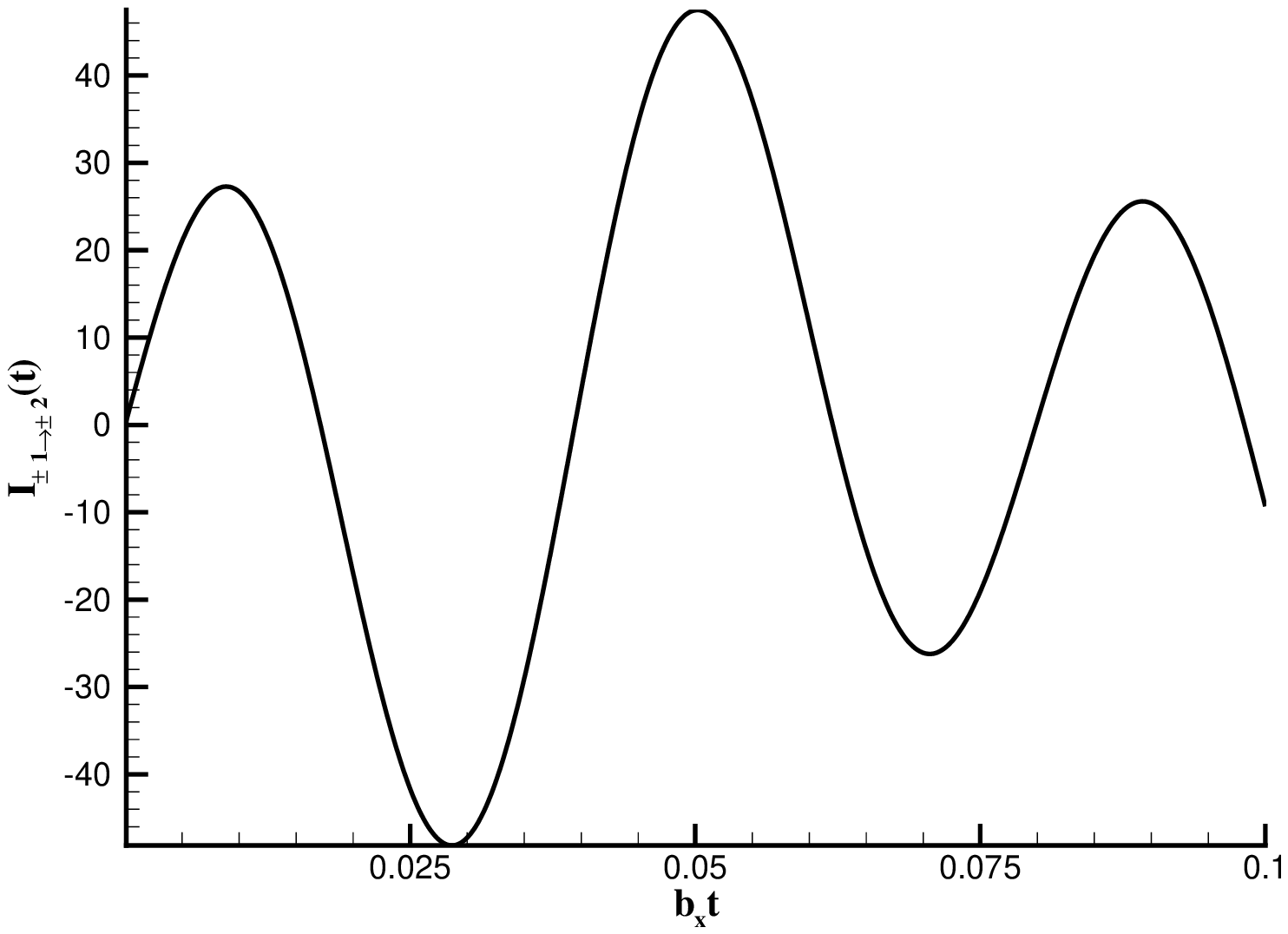}
\caption{ Time evolution of tunneling current $I_{\pm
1\rightarrow\pm 2}(t)$ with an external cosinusoidal magnetic
field, $b_x/\omega=40, b/\omega=8$.}
\end{figure}

\begin{figure}[htbp]
\includegraphics[width=0.7\textwidth,angle=0]{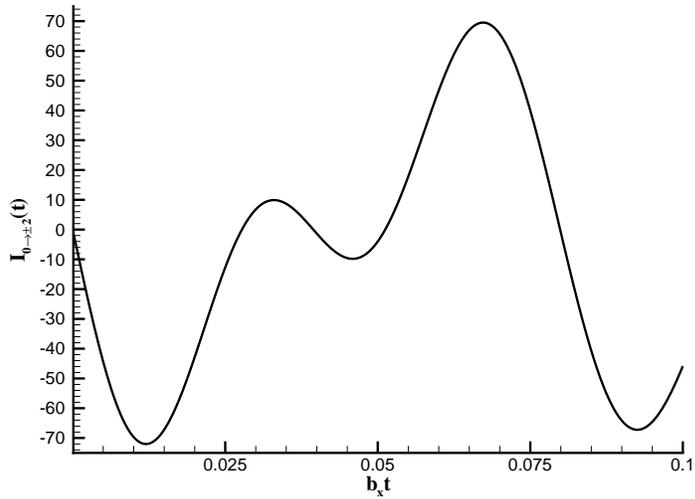}
\caption{Time evolution of tunneling current $I_{0\rightarrow\pm
2}(t)$ with an external cosinusoidal magnetic field,
$b_x/\omega=40, b/\omega=8$.}
\end{figure}

\section{Conclusions}

In this paper, we have studied tunneling effect of the spin-2 Bose
condensate driven by external  magnetic field $b_l(t)=b\cos(\omega
t)$. We find that the population transfers among spin-0 and
spin-$\pm1$, spin-0 and spin-$\pm2$ exhibit the step structure
under the external cosinusoidal magnetic field respectively, but
there do not exist step structure among spin-$\pm1$ and
spin-$\pm2$.  The tunneling current among spin-$\pm1$  and
spin-$\pm2$ may exhibit periodically oscillation behavior, but
among spin-0 and spin-$\pm1$, spin-0 and spin-$\pm2$, the
tunneling currents exhibit irregular oscillation behavior.

\section{Acknowledgments}

This work was supported by the NSF of Shandong Province.

\end{document}